\documentclass{article}

\usepackage{PRIMEarxiv}

\usepackage[%
  natbibapa
]{apacite} 
\usepackage[hyphens]{url}
\usepackage[utf8]{inputenc} 
\usepackage[T1]{fontenc}    
\usepackage[hidelinks]{hyperref}       
\usepackage{url}            
\usepackage{booktabs}       
\usepackage{float}
\usepackage{array}
\usepackage{multirow}
\usepackage{wrapfig}
\usepackage{amsfonts}       
\usepackage{nicefrac}       
\usepackage{microtype}      
\usepackage{lipsum}
\usepackage{fancyhdr}       
\usepackage{graphicx}       
\graphicspath{{media/}}     

\title{Ever heard of ethical AI? Investigating the salience of ethical AI issues among the German population.
}

\author{
  Kimon Kieslich
   \href{https://orcid.org/0000-0002-6305-2997}{\includegraphics[scale=0.06]{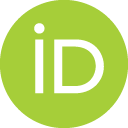}}\\
  Department of Social Sciences \\
  Heinrich Heine University \\
  Düsseldorf, Germany\\
  \texttt{kimon.kieslich@hhu.de} \\
   \And
  Marco Lünich
  \href{https://orcid.org/0000-0002-0553-7291}{\includegraphics[scale=0.06]{orcid.png}}\\
  Department of Social Sciences \\
  Heinrich Heine University \\
  Düsseldorf, Germany\\
  \texttt{marco.luenich@hhu.de} \\
    \And
  Pero Došenović
  \href{https://orcid.org/0000-0002-9515-703X}{\includegraphics[scale=0.06]{orcid.png}}\\
  Department of Social Sciences \\
  Heinrich Heine University \\
  Düsseldorf, Germany\\
  \texttt{pero.dosenovic@hhu.de} \\
}

\begin{document}
\maketitle

\begin{abstract}
Building and implementing ethical AI systems that benefit the whole society is cost-intensive and a multi-faceted task fraught with potential problems. While computer science focuses mostly on the technical questions to mitigate social issues, social science addresses citizens’ perceptions to elucidate social and political demands that influence the societal implementation of AI systems. Thus, in this study, we explore the salience of AI issues in the public with an emphasis on ethical criteria to investigate whether it is likely that ethical AI is actively requested by the population. Between May 2020 and April 2021, we conducted 15 surveys asking the German population about the most important AI-related issues (total of N=14,988 respondents). Our results show that the majority of respondents were not concerned with AI at all. However, it can be seen that general interest in AI and a higher educational level are predictive of some engagement with AI. Among those, who reported having thought about AI, specific applications (e.g., autonomous driving) were by far the most mentioned topics. Ethical issues are voiced only by a small subset of citizens with fairness, accountability, and transparency being the least mentioned ones. These have been identified in several ethical guidelines (including the EU Commission's proposal) as key elements for the development of ethical AI. The salience of ethical issues affects the behavioral intentions of citizens in the way that they 1) tend to avoid AI technology and 2) engage in public discussions about AI. We conclude that the low level of ethical implications may pose a serious problem for the actual implementation of ethical AI for the Common Good and emphasize that those who are presumably most affected by ethical issues of AI are especially unaware of ethical risks. Yet, once ethical AI is top of the mind, there is some potential for activism. 
\end{abstract}


\hypertarget{introduction}{%
\section{Introduction}\label{introduction}}

With the rapid development of AI technology, great threats to society,
some of them even existential, can be expected to materialize and
society has to be prepared to react to those changes
\citep{Hancock.2021}. This cautionary sentiment is shared by other
scholars, albeit with a focus on distinct detrimental consequences
\citep{Bostrom.2016, Floridi.2018}.

Thus, fairness, accountability, and transparency are increasingly
discussed in the scientific community as are other related concepts that
aim for Artificial Intelligence (AI) development for the Common Good.
The Common Good encapsulates the ambition that AI should benefit
everyone or ``be good for all'', not only the ones who invest in or
deploy the technology \citep[p.~45]{Berendt.2019}. While normatively
building ethical AI systems in the interest of everybody---which is
often formulated as a central aim, for example by the EU
\citep{EuropeanCommission.2019}---research has shown that there exist
considerable obstacles and costs on the way to achievement. For
instance, according to \citet{Elzayn.2020} market interests can be in
opposition to developing fair systems as developing such systems
requires additional resources and is cost-intensive. Aside from economic
interests, political interests may not necessarily prioritize ethical AI
systems either \citep{Crawford.2021}. Political actors often rather
argue that AI development must be pushed forward for the sake of
international competition, aiming at certain economical or scientific
goals \citep{Hagendorff.2020}. Concerning the primary aim of
politics to achieve and maintain power or the primary goal of economics
to optimize businesses and grow, these actors are not necessarily aiming
at a broad and society-including discussion about ethical AI which may
oppose their primary goals \citep{Bareis.2021, Hagendorff.2020}.

This paper starts from the assumption that neither academic debates nor
political promises or declarations of intent alone will lead to shaping
AI in the public interest. On the contrary, given the inherent logic of
politics and economics in capitalist society, the claim for ethical AI
will have to be asserted against the interests of economic and political
elites. Regardless of the (inter)national economic and political
contexts in which these interests operate, this can only succeed if a
significant mass of citizens actively requests ethical principles in
their roles as users and consumers of these technologies, or if they
demand credible policies for ethical AI as voters. Hence, as a proposed
safety measure against detrimental consequences of the emerging
technology, one may demand the implementation of AI systems that a
priori fulfill certain ethical criteria. Though, as argued above, the
fulfillment of respective calls for ethical AI does not follow
automatically but must be---at least to some extent---expressed by
societal groups or the broader public in the form of articulated claims
and requests to the political decision-making bodies. Thus, civil
society has to issue convincing incentives for economic or political
actors to invest in AI systems that benefit the Common Good
\citep{Kieslich.2021c}. Consequently, there must be public awareness of
social problems for them to lead to political actions. While we have
some knowledge of how citizens react to the ethical design of AI
\citep{Kieslich.2022, Konig.2021, Shin.2020b, Shin.2021, Shin.2021b, Shin.2021d},
we only have little to no knowledge about the salience of ethical AI
among the public in the first place. Hence, we ask, what issues are top
of citizens' minds and, in particular, how prominent ethical issues
feature among these. We subsequently ask whether the salience of ethical AI
leads to different behavioral intentions.

To shed light on these questions, we first reflect on the current state of
approaches to achieving ethical AI, which is primarily connected to
ethical guidelines. We highlight the shortcomings of the current
approaches to ethical AI, emphasizing their conceptual weaknesses as
well as the non-binding character of proposed ethical guidelines. We
infer that public pressure must be exerted to achieve ethical AI for the
Common Good. As the issues of emerging technologies like AI are
mostly raised by the media, we then discuss the current literature on
published opinion on AI. We especially reflect on the topical structure of
the media reporting, i.e.~the prevalence of ethical topics, as well as
the presence of different actors. Afterward, we review studies about
public opinion on AI and highlight that no study so far has researched
citizens' salience of ethical AI without confronting respondents with
ethical problems of AI beforehand.

In the empirical part of the paper, we first explore the extent to which
German citizens are concerned with ethical issues of AI and compare that
to the salience of other phenomena and consequences of AI, for instance,
technical functionalities or specific applications of AI. Second, we
show which factors (sociodemographic as well as interest in AI) are
linked to a higher likelihood of being concerned with ethical issues.
Thereby, we especially look for differences in societal groups and
explore whether groups of lower social status---those groups, that are
arguably most endangered by contemporary AI technologies---are aware of
those issues. Or, in other words, whether salience of ethical issues
with AI is a discourse, which primarily bears fruit with people of
higher societal status. In a third step, we investigate whether the issue
salience of ethical AI has an impact on behavioral intentions towards
AI, namely the intention to avoid AI and the intention to engage in
discussions about AI.

We point out theoretical and practical implications in light of the low
salience of ethical AI with the German public and the differences among
societal groups in their respective salience levels. We further discuss
the potential of the salience of ethical AI for the articulation of ethical
demands to those, who are in charge of the development and deployment of AI
systems. In this paper, we show that ethical AI discourse needs to be
taken into all levels of society.

\hypertarget{pathways-to-ethical-ai}{%
\section{Pathways to Ethical AI}\label{pathways-to-ethical-ai}}

\hypertarget{ethical-ai-guidelines}{%
\subsection{Ethical AI Guidelines}\label{ethical-ai-guidelines}}

Against the backdrop of ethical problems in AI development, some
approaches were developed to tackle those issues that emerge with
further development and implementation of AI in society.
\citet{Jobin.2019} collected and analyzed corresponding ethical AI
guidelines from around the globe. Interestingly, many ethical guidelines
were proposed by private companies or political institutions (e.g., the
EU-Commission proposed a framework for the development of ethical AI
\citep{EuropeanCommission.2019}), but also by academia and research
institutions such as the Association for Computing Machinery (ACM)
\citep{AssociationforComputingMachinery.2018, Jobin.2019}. However, many
of these guidelines do not address the broad public or civil society
and instead target specific stakeholder groups. Thus, they focus only on
the \emph{good} of a subset of society or speak to those who are in
direct contact with the technology. Guidelines that address the Common
Good \citep{Berendt.2019} rarely occur. AI for the Common Good can be
defined as an approach to developing AI systems that take many different
social groups into account and not only those, who ``directly pay for
the development or use of this AI'' \citep[p.48]{Berendt.2019}.
Consequently, ethical AI guidelines can not necessarily be equated with
AI that benefits the whole society or that may recognize and include
multiple societal perspectives at all. Looking at the sources and target
groups of the ethical guidelines summarized by \citet{Jobin.2019}, only
a small subset aims for AI for the Common Good or the
\emph{broad public}, respectively. However, that does not imply that the
other ethical guidelines exclude these aims, but they simply do not
mention or prioritize them as the main goal to satisfy respective
stakeholders.

But what about the effectiveness of ethical guidelines? Given the
different aims of stakeholder groups, it is debatable to what extent
these actors want a broad discussion about ethical AI or only treat
guidelines primarily for communicative purposes with other stakeholders.
Out of their logic of achieving or maintaining power (political
perspective) or increasing monetary profit (economic perspective), one
can argue that most political or economic actors are not aiming for a
broad and society-including discussion about ethical AI, as this can be
opposed to their primary aims.

\citet{Hagendorff.2020} analyzed 22 ethical guidelines from companies as
well as research institutes and political institutions and attested the
current state of AI ethics a bad condition: ``Currently, AI ethics is
failing in many cases. Ethics lacks a reinforcement mechanism.
Deviations from the various codes of ethics have no consequences.''
(p.113). He argues that current ethical guidelines mainly serve as a
``marketing strategy'' (p.113) and that ethics are seen as an add-on
rather than a crucial step integrated into AI development. This is
especially true in economic terms that oftentimes are in contrast to---and in the end override---commitment to ethical principles in AI
design leading to a violation of ethical principles. Further, he
highlights the non-binding character of these guidelines, which have the
aim to avoid state regulation. In addition, focusing on the political
approaches of the US, the EU, and the UK for AI for a ``Good Society'',
\citet{Cath.2018} report that, although issues are adequately addressed,
the reports fall short in providing clear political perspectives on how
to tackle these issues. Thus, also in the context of political
guidelines, the effectiveness of those guidelines is questionable.
Turning towards the practical side of developers' adherence to ethical
guidelines, \citet{McNamara.2018} surveyed software
engineers and report that reading ethical guidelines did not change their
reported working habits. Thus, the prevalent top-down approach to
ethical guidelines appears to have only a limited impact on the ethical
AI development.

\hypertarget{the-role-of-the-public-in-implementing-ethical-ai}{%
\subsection{The Role of the Public in Implementing Ethical
AI}\label{the-role-of-the-public-in-implementing-ethical-ai}}

Accordingly, the expression of AI ethics guidelines is not a sufficient
requirement to achieve the goal of an AI for the Common Good. From a
civil society perspective, therefore, the question is how to commit
relevant stakeholders to the goals of an AI for the Common Good. How can
`bottom-up' political pressure be created that encourages those
responsible to commit themselves and each other to the observance of
corresponding guidelines and to stick to their commitments? Here,
especially in pluralistically organized democratic societies, the focus
must inevitably turn to the public sphere as the place where the
electorate informs itself about political issues and forms an opinion.
Irrespective of which specific conception of democracy and its public
sphere one follows---for example, deliberative, participatory, and
representative conceptions of democracy
\citep{Ferree.2002, Ferree.2002b}---the public and the media have
important roles to play in the processes of political negotiation and
legitimization \citep{Beetham.2013, Jungherr.2019}. From a normative
perspective, the media is often expected to serve as a watchdog
\citep{Norris.2014b, Bennett.2005}, while the public---represented in
part by advocacy groups---voices support for or disapproves of the
implementation of technology. This results in the widespread continuous
monitoring and analysis of public opinion that assesses citizens'
attitudes and understanding of technology
\citep{Besley.2013c, Miller.2004}.

Media coverage is especially relevant in the case of emerging
technologies, since most people do not have personal experience with
such technology or are not even aware of it, respectively. Thus, media
present those topics to a broad public and have the potential to shape
public opinion in setting a frame of reference for what and how to think
about technology, or, at least, what relevant and powerful actors
think about AI \citep{Nisbet.2002, Scheufele.2005, Metag.2014}. With
that, the audience learns about potential opportunities and risks of a
technology as well as future pathways for dealing with the emerging
sociotechnical systems. Thus, citizens' opinion toward emerging
technology is affected by the way the media reports on it. Accordingly,
it is important to understand how the media report on AI, especially
concerning the topical structure and the presence of different social
groups and actors (e.g., civil society, political actors, economic
actors).

Taken together, public attention concerning ethical issues of AI thus
may result in substantial pressure on the actual implementation of
ethical AI, as affected groups and concerned citizens, reject
sociotechnical AI systems altogether (`exit') or protest their
implementation (`voice') \citep{Hirschman.1970, Kieslich.2021c}. For
instance, in the UK, students protested against an algorithmic
decision-making (ADM) system that autonomously graded exams, which lead
to a withdrawal of the respective system \citep{Kelly.2021}. In France,
a public debate about a university admission system also ended in
abolition of the system \citep{Wenzelburger.2021}. Concerning the use of
an algorithm to allocate Covid-19 vaccination in Stanford hospital,
medical staff heavily protested against the decisions of the algorithm,
as front workers were not given priority \citep{Guo.2020}. Additionally,
journalists and activists also raised concerns about the implementation
of unethical AI systems, for example in Sweden in the use case of a
social benefit application and in Germany in the use case of face
recognition in public places \citep{AlgorithmWatch.2020}. Additional
studies suggest that the perceptions of unfair treatment by ADM systems
can lead to the rejection of such technology and of those who apply it,
respectively \citep{Marcinkowski.2020}; furthermore, \citet{Lunich.2022}
noted that a lack of trust in ADM systems leads to them being perceived
as illegitimate. However, Algorithm Watch estimates the awareness of
ethical problems with AI as not being overly high: ``Protests in the UK
and elsewhere, together with high-profile scandals based on ADM systems,
have certainly raised awareness of both the risks and opportunities of
automating society. But while on the rise, this awareness is still in
its early stages in many countries.'' \citep[p.9]{AlgorithmWatch.2020}

When analyzing the given examples, two reasons for reluctance towards AI
systems are apparent. First, reactance towards AI is tied to a
\emph{specific use case} and articulated by those who are at risk of
being treated negatively (e.g., students or medical staff in the
examples mentioned above). These stakeholder groups organized themselves
and fueled widespread protests against the use of the specific AI
application. Second, \emph{media reporting} led to public attention
about ethical issues with some---again specific---AI systems. Media
reporting may not lead to immediate protest behavior in the public, yet,
it can damage the public image of developers or suppliers of AI
technology. Hence, they may proactively work on ways to resolve ethical
issues before widespread protests can emerge.

Thus, analyzing public perceptions of ethical issues of AI can indicate
how strong the public sphere is concerned with AI implementation and
what public demands and objections may follow from social debates around
the ethics of AI. In this study, we do not focus on a specific
application but on the general attitudes towards AI and its ethics.

\hypertarget{published-and-public-opinion-on-ai}{%
\section{Published and Public Opinion on
AI}\label{published-and-public-opinion-on-ai}}

Focusing on the inclusion of public perceptions in the debate, we
first need to address the extent of AI's ethical issues that are
discussed in the sphere of the media. We then turn to the analysis of
public opinion concerning AI and AI ethics to document whether issues
from media discourse are also reflected within public opinion.

\hypertarget{published-opinion-on-ai}{%
\subsection{Published Opinion on AI}\label{published-opinion-on-ai}}

In recent years, a considerable amount of research has focused on the
analysis of news reporting on AI
\citep{Brennen.2018, Chuan.2019, Fast.2017, Fischer.2021, Ouchchy.2020, Sun.2020, Vergeer.2020, Zeng.2020}.
According to the studies of \citet{Fast.2017}, \citet{Sun.2020} and
\citet{Chuan.2019} for the US, \citet{Fischer.2021} for the German and
\citet{Vergeer.2020} for the Dutch press, media reporting about AI
sharply increased in recent years, leading to a steady presence of AI in
news reporting. Thereby, several studies pointed out that media
reporting about AI was more frequent by right-leaning, respectively
conservative outlets \citep{Brennen.2018, Fischer.2021}. Further,
\citet{Vergeer.2020} found for the Dutch press that national newspapers
and tabloid newspapers report more on AI than the regional press.

Focusing on the topical structure of news reporting about AI, several
studies across different countries found that economic topics dominate
the content of the news
\citep{Brennen.2018, Zeng.2020, Fischer.2021, Chuan.2019}. Accordingly,
it is economic actors that dominate the media coverage of AI across the
globe; oftentimes with a special focus on so-called `tech giants' (i.e.,
corporations that dominate the space of information and communication
technology)
\citep{Brennen.2018, Sun.2020, Fischer.2021, Vergeer.2020, Zeng.2020}.
On the other hand, civil society actors or NGOs are rarely featured in
news reporting about AI
\citep{Brennen.2018, Sun.2020, Fischer.2021, Vergeer.2020, Zeng.2020}.
For example, in German \citep{Fischer.2021} and UK media reporting
\citep{Brennen.2018} only four percent of the articles analyzed cited
actors from civil society. Moreover, several studies indicated a
positive \citep{Fischer.2021, Fast.2017, Zeng.2020} or ambivalent
sentiment regarding AI in news reporting
\citep{Chuan.2019, Vergeer.2020}, while none of the studies reported
that AI is portrayed as particularly negative. Positive sentiment is
often connected to an optimistic turn towards AI
\citep{Fast.2017} or based on a positive wording about AI
\citep{Vergeer.2020, Zeng.2020}. Additionally, two studies
explicitly focused on the occurrence of opportunities and risks in media
reporting. \citet{Fischer.2021} report for the German context that the
number of articles analyzed that outlined opportunities of AI was more
than double as high as the number of articles that reported on risks of
AI. Opportunities were mostly associated with economic progress and
efficiency and risks, in contrast, with a lack of AI competence within
the general population (i.e., not ethical AI). For the US media
coverage, \citet{Chuan.2019} report that there is an emphasis on
opportunities compared to risks about AI as well.

Looking at news coverage about ethical issues of AI, \citet{Chuan.2019}
reported that the topic of AI ethics received increased attention by the
US-American press from 2018 on. However, in comparison with other
perspectives on AI, it is still a niche topic. \citet{Ouchchy.2020}
specifically analyzed news articles about ethical issues of AI using an
English-language sample of various media outlets. They found that
prejudices reflected in data, privacy, and transparency were the most
common issues in the news articles. Furthermore, they report that when
specific applications were discussed in terms of AI ethics,
\emph{autonomous driving} was the most mentioned application followed by
the use of AI in the military and healthcare domain. Overall, the
discussion about AI ethics was neither particularly critical nor
enthusiastic. For the German media landscape, \citet{Fischer.2021}
report that, if risks of AI were mentioned in the news, especially
accountability and transparency issues were featured.

Summing up the literature on news coverage of AI, some similarities
emerge on a global level. The media discourse is led by economic and
political actors, and topics that highlight the positive impact of AI.
Ethics is a niche topic in media reporting about AI. These findings also
apply to Germany, where the present study was conducted.

The empirical results on published opinion on AI may be partly explained
by the dominance of conservative and business-friendly publishers
\citep{McChesney.2008, Wasko.2011, Nichols.2020} of news stories about
AI among the press. This explains the observation that especially those
actors are most present in the mass media within the public sphere who
utilize ethical guidelines as a marketing device or have a vested
interest in an unregulated approach to AI, while actors who actively
strive for implementation of ethical AI for the Common Good are
fairly underrepresented. In absence of personal experience with AI, from
the perspective of communication science, the underlying assumption is
``that the media crucially influence audience attitudes towards emerging
technologies'' \citep[p.465]{Metag.2014}. For the salience of AI issues
among the public this may mean the following: On the one hand, news
media weigh some topics over others and the public gets to know a rather
one-sided picture of the technology with a strong emphasis on economic
opportunities. On the other hand, when there is no media interest in
reporting on ethical AI, then salience of AI-related ethical issues
among the public seems to be highly unlikely, too.

\hypertarget{public-opinion-on-ai}{%
\subsection{Public Opinion on AI}\label{public-opinion-on-ai}}

Following increasing academic interest in media reporting on
AI, several studies have recently explored public opinion
on AI.

Research on public opinion on AI found that, while many people have
heard about the term, usually, only some of the surveyed people suggest
to have an idea of what AI is and what AI entails, or claim to be
somewhat knowledgeable about AI, respectively \citep{Cave.2019, Selwyn.2021}.
Thereby, AI is mostly associated with technical terms, foremost robots
or computers \citep{Cave.2019, Selwyn.2021, Kelley.2021}. All in all,
when surveyed, current AI technology is perceived rather positively in
various countries \citep{Kelley.2021, Fietta.2021, Zhang.2019}. In
addition many studies conducted around the globe, highlight
sociodemographic differences in the evaluation of AI: Male, younger, and
well-educated respondents, people with higher income, and people, who
consider themselves to possess much knowledge about AI are more likely
to expect that opportunities materialize with the advent of AI
\citep{Carradore.2021, Fietta.2021, Zhang.2019}. Moreover,
\citet{Kelley.2021} report country differences in the strength of
positive sentiment towards AI, whereby residents of non-Western
countries (e.g., Brazil, and South Korea) are more enthusiastic about AI
than residents of Western countries (e.g., USA, France).

Some studies investigated public opinion towards selected ethical
aspects of AI using closed-ended questions. Utilizing a representative
German sample, \citet{Kieslich.2020} asked for risk perceptions of some
particular ethical challenges that AI might pose in terms of injustice
and discrimination. On the one hand, the authors report that uncertain
accountability, as well as loss of control, were perceived as major
ethical challenges of AI implementation by the public. On the other
hand, potential systematic discrimination of social groups or injustice
via AI were only articulated by a subset of the German population when
inquired. The results also suggest that a significant portion of
respondents (17\% and 16\%, respectively) did not answer or indicated
that they did not know whether the latter issues pose a social risk in
contrast to 5\%-9\% missing answers for other ethical issues. This
indicates that fairness and discrimination issues may not be salient or
even conceivable for a fair share of the German population. Another
study by \citet{Kieslich.2022} researched preferences for the ethical
design of an AI system in the use case of tax-fraud detection. They
report that accountability was perceived as the most important ethical
criterion among the respondents, while machine autonomy was perceived as
the least important. However, the researchers also report that the
different ethical AI principles were perceived as roughly equally
important, even though the respondents show different preference
patterns among each other. Nearly one-quarter of the respondents cared
only marginally about the ethical design of AI systems, while 32\% were
highly concerned about AI ethics. Moreover, the results suggest that
those that were not concerned with ethical AI were predominantly older,
less educated, and less interested in AI than respondents that were
concerned about the ethical design of AI systems. In summary, the German
public expresses some concerns about AI ethics, but the ethical concerns
vary between different segments of the population
\citep{Kieslich.2020, Kieslich.2022}.

Researching the perceived importance of ethical requirements proposed by
the EU, \citet{Choung.2022} showed for a US sample that, again,
accountability was perceived as most important. As in the study by
\citet{Kieslich.2022}, they also report that all ethical requirements are
evaluated equally important. \citet{Ikkatai.2022} surveyed
Japanese citizens to evaluate public attitudes towards AI ethics.
Testing for four different scenarios, they show that public attitudes
toward the importance of ethical AI design are context-dependent, with
AI use in the military being the most ethically problematic one. AI use in
a military context was also the scenario most citizens disagreed with.
\citet{Ikkatai.2022} also showed that age was significantly correlated
with attitudes towards AI ethics in all scenarios; moreover, gender,
interest, and AI literacy were significantly correlated to attitudes
toward AI ethics in some scenarios.

All in all, when it comes to public opinion concerning AI ethics,
there is only limited evidence towards \emph{general salience} of
ethical issues connected with AI. On the one hand, the research at hand
often addresses these issues as a rather broad category within the topic
of AI and it is not clear what specific ideas citizens have when it
comes to the ethics of AI. On the other hand, research that directly
gathers \emph{opinions} on specific ethical aspects often does not
capture the general \emph{salience} of ethical issues of AI with
citizens. In asking directly about opinions of ethical aspects of AI,
the question itself makes ethical aspects salient; also to those, who
would otherwise not indicate having thoughts about ethical AI, to begin
with. Therefore, we do not know much about the actual salience of
ethical AI and its consequences on intended behaviors among citizens.
Thus, in this study, we bridge the literature strands of broad
associations with AI and concrete opinions on ethical AI in
investigating the general salience of ethical AI with the public.

\hypertarget{hypotheses-and-research-questions}{%
\section{Hypotheses and Research
Questions}\label{hypotheses-and-research-questions}}

In the present study, we build on the existing literature and dig deeper
into public opinion on AI with a focus on ethical AI. Our study was
conducted in Germany, which-politically-follows the Western,
respectively European approach to AI. That is, the German government
aims for broad funding of economical and scientific progress in AI
while acknowledging ethical issues \citep{DieBundesregierung.2018}.
Thus, many AI projects were funded by the German government in recent
years. However, as a small inquiry by the Left Party in the German
Bundestag revealed, that only a few AI systems were checked with an AI
risk analysis before their implementation
\citep{DieBundesregierung.2022}. Regarding the state of AI usage in the
economic sector in Germany, 5.8 percent of all companies used AI methods
in 2019 \citep{BundesministeriumfurWirtschaftundEnergie.2020}, while,
according to a study conducted in 2021, 30 percent of German companies
indicated that they are thinking about implementing AI
\citep{Bitkom.2021}. Hence, AI plays a continually growing role-at
least in politics as well as in the economy.

While former public opinion studies utilized open-answered questions
with a focus on knowledge about AI, we are interested in the issues that
are present in the minds of citizens---if any. Unlike other studies, we
do not present respondents with a given set of different future
perspectives or ethical issues but gather information if citizens link
AI to ethical aspects on their own. As AI is still a relatively new
technology and news reporting of AI is---albeit increasing ---still a
topic that is not on the top of the news, we first explore how many
respondents are preoccupied with AI topics at all. In a second step, we
are interested in the different topics citizens associate with AI. As
discussed earlier, especially economic and political actors, which are most prominent
in media reporting, are not interested in a wide discussion about the
social impact of AI. Consequently, we suppose that issue salience of AI
ethics is not very prominent among German citizens. Another reason for
that is that in the German context, unlike in the UK, no major scandal
concerning AI emerged in the past. Yet, we do not know the variety and
the emphasis of different AI issues on behalf of citizens. Thus, we set
up RQ1.

\emph{RQ1}: What topics are present in the German population with regard
to AI?

Further, prior studies suggest that several individual characteristics
may influence public opinion on AI such as interest in or knowledge of
AI as well as sociodemographic aspects like gender, age, educational
level or socioeconomic status
\citep{Kieslich.2022, Ikkatai.2022, Carradore.2021, Fietta.2021, Zhang.2019}.
Given the one-sided media reporting about AI and the focus on specific
stakeholders by economic and political actors, we suppose that AI can be
considered as a topic of the elite. Thus, we hypothesize:

\emph{H1a}: Salience of AI issues is higher among citizens with a higher
level of education.

\emph{H1b}: Salience of AI issues is higher among citizens with a higher
socioeconomic status.

\emph{H1c}: Salience of AI issues is higher among citizens with a higher
general interest in AI.

Additionally, we explore which socio-economic characteristics and
interest in AI influence if someone is concerned with ethical AI in
particular. In the context of ethical aspects of AI, the questions of which
sociodemographic factors, as well as interest in AI, influence the issue
salience of ethical AI, become particularly relevant, since AI
implementation can have serious consequences for different societal
groups. For example, it was found that AI systems treated women
\citep{Tambe.2019} or persons with a low socioeconomic status
\citep{Pandey.2021} worse than others. Additionally, \citet{Frey.2017}
assumed that the jobs of people with a lower educational degree are more at
risk of being automated. Thus, we explore if those who are probably more
endangered by the implementation of AI systems are aware of
those ethical issues. Though, due to a lack of media reporting on
ethical aspects and the dominance of economic topics, affected groups
might not be concerned with ethical issues of AI. Hence, we ask:

\emph{RQ2}: Do educational level, socioeconomic status, and interest in
AI have an effect on salience of \emph{ethical} AI issues?

Further, we investigate if different topics are associated in context
with each other. We suppose that ethical issues are mostly connected to
an anchor example. Thus, ethical issues emerge when discussed in light
of a certain usage of AI. For example, \citet{Ouchchy.2020} showed that
ethical aspects of AI were repeatedly reported on in the context of
specific applications like autonomous driving. Also, the cases were
protest was articulated by the population or by civil society actors
were tied to a specific example, e.g.~education admission systems
\citep{Kelly.2021} or vaccine allocation \citep{Lunich.2022}. As the
literature does not suggest, \emph{which} use cases are currently on the
top of the head of citizens or especially suitable for a public debate
about ethical issues of AI, we formulated H2 rather broad:

\emph{H2}: People who are concerned with ethical AI commonly connect AI
to specific applications.

Lastly, we investigate if the salience of ethical AI influences
behavioral intentions. We have outlined that ethical AI has to be
demanded by the public to be put into practice by companies or
politicians. For companies, considering citizens' demands is necessary in
the sense of economic profit. If citizens hesitate to use AI systems
because of ethical considerations or raise protests against an AI tool,
it would damage the profit companies make. Hence, public pressure can
presumably force economic actors to follow ethical guidelines.
Likewise, political actors can be influenced by public opinion as well.
If citizens demand ethical AI systems, political actors could listen to
those voices and set up binding guidelines for the development and
implementation of AI systems. However, for that to happen, the salience of
ethical AI needs to influence citizens' behavior in a way that leads
them to reluctance toward AI. Given the use cases, where ethical flaws
in AI systems lead to public outrage, we presume that the salience of
ethical AI among the public affects AI avoidance intention,
respectively the intention to engage in public discussions about AI.

\emph{H3a}: Salience of ethical issues of AI positively affects intended
AI avoidance.

\emph{H3b}: Salience of ethical issues of AI positively affects the
intention to engage in public discussions about AI.

\hypertarget{method}{%
\section{Method}\label{method}}

\hypertarget{data-collection-sample}{%
\subsection{Data Collection \& Sample}\label{data-collection-sample}}

The study is part of the long-term monitoring of public opinion on AI in
Germany. In fifteen surveys, about 1,000 respondents were asked about
interest, attitudes, and behavioral intentions regarding AI. The monthly
surveys were conducted between May 2020 and April 2021 by the market
research institute forsa GmbH as part of their Omninet omnibus panel.
That is, the questions on AI were asked alongside
other changing topics. The advantage of this procedure is to minimize
self-selection, as respondents do not choose questionnaires by interest
in specific subjects. This allows a more realistic and rather
representative picture of the population.

The Omninet panel consists of 75,000 panelists who are recruited via
random sampling of phone numbers. The panel is representative of the
German population with internet access aged 14 and older for the
personal respondents characteristics age, gender, and regional place of
residence. Our questions were only asked those panelists who are 18
years and older.

All 15 samples included in this study were randomly selected from the
Omninet panel. After data cleaning, the dataset contains 14,988
respondents. Of these, 51.6\% are self-identifying female and the
average age is 51 (\emph{SD} = 15.96) years. This corresponds very well
to the distribution in the German population aged 18 and older with
internet access \citep{AGOF.2020}. 60.2 percent of our sample have at
least a high school degree, higher education being slightly
overrepresented \citep{Dosenovic.2021b}.

\hypertarget{measurement}{%
\subsection{Measurement}\label{measurement}}

\emph{Salience of AI-related issues.} As for the dependent variable, we are
interested in whether people are concerned with AI and, if so, what
these issues are. Following agenda-setting research
\citep{McCombs.1972}, we are looking for a measurement of the
public agenda---but this time precisely on the subject of AI. Two
conditions were important to us: First, respondents should be able to
say that no issue concerns them at all. Second, we did not want to ask
about problems by default, as was often done in agenda-setting research
\citep{Smith.1980}. Thus, we asked ``If you think about recent
times, which issues related to artificial intelligence have been of most
concern to you personally?'' Respondents were asked to provide up to
three answers to this question in an open text box.

After conducting the fourth survey, the codebook was developed in an
iterative process. That is, the two coders and one of the authors
randomly drew several samples of responses from the data set. They coded
all the same data and discussed their results afterward. Each of the
possible three responses was assigned to only one code. If two (or more)
codes were deemed applicable, the coding that described AI in more
detail or on which emphasis was placed was selected. For example, if an
answer was ``job loss due to automation'', ``job loss'' was chosen as
code as it depicts the consequence of automation. Analogously, ``ethical
issues due to autonomous driving'' was coded as ``ethics''. After the
first coding process, additional samples were drawn and previously
formulated codes were used to categorize the new data until there was no
new code assigned. This resulted in four major categories (technical
functionality; AI in use / AI applications; ethical issues about AI;
other issues about AI) with a total of 49 subcodes. A new sample of 150
responses was then coded by the two coders for a reliability test.
Krippendorf's alpha \citep{Hayes.2007} is rated as good at the value of
\(\alpha_{K}\) = .82 {[}Bootstrap with 10,000 samples; CI 95\%: .72;
.90{]}. For the category \emph{ethical issues about AI}, we adapted the
principles outlined by \citet{Jobin.2019}. We added the more specific
ethical problems \emph{surveillance}\footnote{Surveillance was assigned
  to the ``ethical issues'' category, since most answers referred to
  surveillance in the dystopian sense of mass surveillance.},
\emph{manipulation} as well as a general subcategory for \emph{ethics}
to the list of ethical criteria, as they emerged as subcategories during
the coding process. The codebook with the occurrence of all subcodes in the
sample can be found in Appendix A.

\emph{Interest in AI.} Interest in AI was measured with the following
four items on a five-point Likert scale: ``I follow AI processes with
great curiosity.'', ``I am very interested in AI in general.'', ``I read
articles about AI with great attention.'', ``I watch or listen to
articles about AI with great interest.'' The index suggests high
internal consistency (\emph{M} = 2.49; \emph{SD} = 0.97; \(\alpha_{C}\)
= .94).

\emph{Behavioral Intentions}. We were interested in how the salience of
specific AI-related issues is associated with the intention to avoid AI or
to engage in public discussions about AI, respectively. Both were
measured as single items on a five-point Likert scale: ``I will stay
away from artificial intelligence wherever possible.'' (\emph{M} = 2.43;
\emph{SD} = 1.18) and ``I will express my opinion in public discussions
about artificial intelligence.'' (\emph{M} = 2.67; \emph{SD} = 1.26).

\emph{Controls}. In the results part, we utilized age (in years), gender
(0 = male, 1 = female), educational level (1 = primary degree, 2 =
secondary degree, 3 = tertiary degree), and socioeconomic status (0 =
household income above the median, 1 = household income below median) as
control variables.

\hypertarget{results}{%
\section{Results}\label{results}}

All in all, 6,221 (41.5\%) respondents indicated that they were recently
concerned with AI and, thus, gave at least one answer to the question at
hand. 2,090 (13.9\%) respondents gave at least two answers and 1,030
(6.9\%) respondents gave three answers. In contrast, 7699 (51.4\%)
respondents were not concerned with AI, while 302 (2.0\%) respondents
stated that they do not know the term artificial intelligence, and
further 766 (5.1\%) preferred not to answer this question.

First, we were interested in the specific AI issues the German public
was concerned with (RQ1). In general, specific application domains were
the most frequent response. 4,122 (66.3\%) named at least one application
or application domain in the open-ended answers. In contrast to this, AI
functionalities were only mentioned by 906 (14.6\%), AI issues by 1,274
(20.5\%), and AI ethics by 943 (15.2\%) of the respondents. Thus, the
German public is mostly concerned with specific use cases of AI rather
than issues, functionalities, or ethical issues

To dig deeper into the specific topics that the German public
is concerned with, we also investigated the subcodes of the categories.
Figure 1 shows the most common issues with AI and groups them according
to their dimensions (functionalities, applications, ethical issues,
and other AI issues). For better readability of the figure, we only depicted
those issues that at least 100 respondents mentioned.

\begin{figure}
\includegraphics[width=1\linewidth]{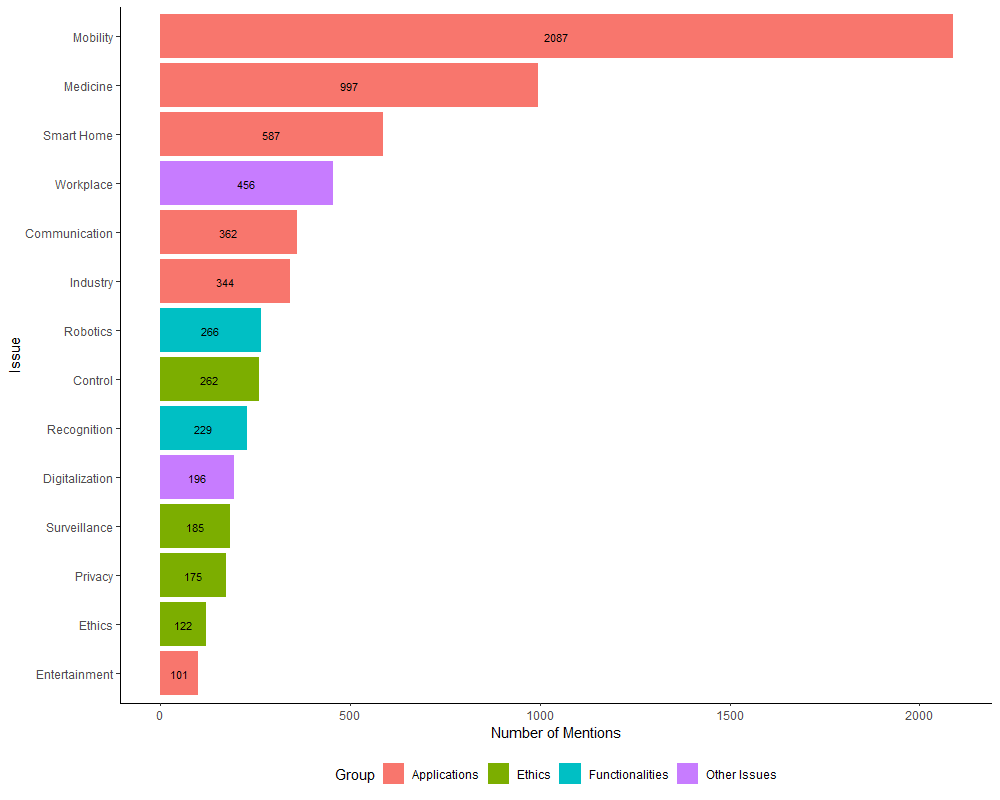} \caption{Concerns about AI among the German Population}\label{fig:unnamed-chunk-2}
\end{figure}

The result shows that the most salient issue was one specific
application domain, namely autonomous driving or mobility. It was
mentioned by nearly one-third of the respondents who were concerned with
AI at all. Also, other application domains such as medicine or smart home
applications were frequently mentioned. Concerning AI issues, the most
frequent mentions were connected to workplace issues such as job
opportunities or job losses due to AI. Regarding technical issues, AI
was frequently connected to robotics, which is in line with the study of
\citet{Selwyn.2021}. Only a small portion of respondents name ethical
aspects as the most salient issues.

Nevertheless, we were interested in the gradations of ethical AI issues.
Thus, we also counted the mentions of all ethical issue subcodes. Figure
2 shows the result. Ethical issues are most connected with indications
about control, i.e.~if AI can or needs to be controlled. Another, more
specific ethical concern is the possibility of surveillance through AI
technology, which was the second most ethical issue raised by the
respondents. Remarkably, terms that are connected with the FAccT
dimensions are the ethical dimensions that are least mentioned by the
respondents. Thus, there is practically almost no salience of fairness,
accountability, and transparency issues among the German public.

\begin{figure}
\includegraphics[width=1\linewidth]{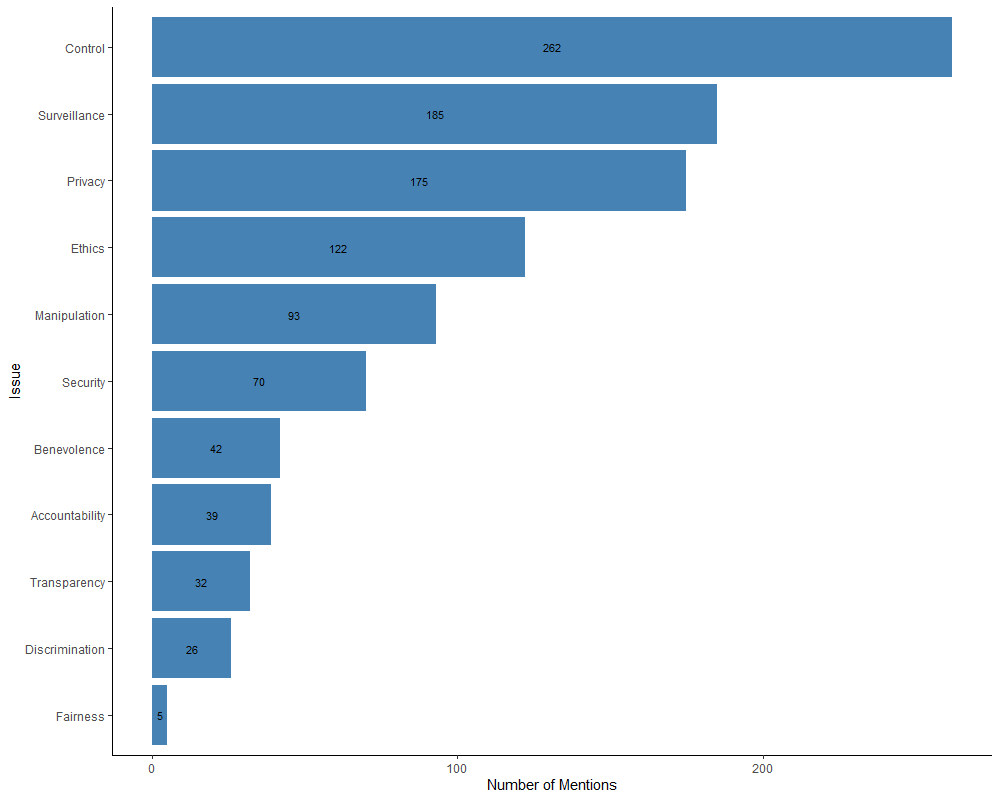} \caption{Concerns about Ethical Issues among the German Population}\label{fig:unnamed-chunk-3}
\end{figure}

In the next step, we research the hypothesized effects of educational
level, socioeconomic status, and interest in AI on the salience of AI in
general (H1a-H1c). We included the age and gender of respondents as control
variables in the model.

\begin{table}

\caption{\label{tab:unnamed-chunk-4}Logistic Regression on the Salience of AI Issues}
\centering
\begin{tabular}[t]{lllllll}
\toprule
\multicolumn{3}{c}{ } & \multicolumn{3}{c}{95\% CI for Odds Ratio} & \multicolumn{1}{c}{ } \\
\cmidrule(l{3pt}r{3pt}){4-6}
  & b & SE & Lower & Odds Ratio & Upper & p\\
\midrule
Intercept & -4.458 & 0.168 & 0.008 & 0.012 & 0.016 & 0.000\\
Age & -0.011 & 0.001 & 0.987 & 0.990 & 0.992 & 0.000\\
SE Status (1=Below Median) & -0.067 & 0.045 & 0.856 & 0.935 & 1.021 & 0.134\\
Gender (1=Female) & -0.029 & 0.044 & 0.891 & 0.971 & 1.059 & 0.509\\
Educational Level & 0.532 & 0.038 & 1.580 & 1.702 & 1.834 & 0.000\\
Interest in AI & 1.342 & 0.028 & 3.628 & 3.827 & 4.041 & 0.000\\
\bottomrule
\multicolumn{7}{l}{\rule{0pt}{1em}\textit{Note: }}\\
\multicolumn{7}{l}{\rule{0pt}{1em}$N$=12,678, Nagelkerke $R^2$=.372, Model $\chi^2$(5)=4,111.5, p<.01}\\
\end{tabular}
\end{table}

By calculating a logistic regression model on the issue salience of AI, we
find that higher educational level, being younger, and being more
interested in AI in general, seeking information about AI,
respectively, predict whether AI is on the personal agenda, as indicated
by expressing at least one AI issue (Table 1). No significant effect
could be found for socioeconomic status. The model explains 37.2 percent
of variance with interest in AI being the strongest predictor. Thus, H1a
and H1c are supported, while H1b needs to be rejected.

In the next step, we investigated the effects of educational level,
socioeconomic status, and interest in AI on the salience of ethical issues.
Again, we conducted a logistic regression on the dependent variable
\emph{salience of ethical AI issues}. Thereby, we only calculated the
model for those respondents who were concerned about at least one AI
issue. Thus, we explore if specific respondents' characteristics lead to
a higher salience of ethical issues. We controlled the effects for
respondents' age and gender (Table 2).

\begin{table}

\caption{\label{tab:unnamed-chunk-5}Logistic Regression on the Salience of Ethical AI Issues}
\centering
\begin{tabular}[t]{lllllll}
\toprule
\multicolumn{3}{c}{ } & \multicolumn{3}{c}{95\% CI for Odds Ratio} & \multicolumn{1}{c}{ } \\
\cmidrule(l{3pt}r{3pt}){4-6}
  & b & SE & Lower & Odds Ratio & Upper & p\\
\midrule
Intercept & -3.107 & 0.319 & 0.024 & 0.045 & 0.083 & 0.000\\
Age & 0.008 & 0.003 & 1.003 & 1.008 & 1.013 & 0.003\\
SE Status (1=Below Median) & 0.198 & 0.080 & 1.041 & 1.219 & 1.425 & 0.013\\
Gender (1=Female) & 0.091 & 0.079 & 0.937 & 1.096 & 1.280 & 0.250\\
Educational Level & 0.183 & 0.075 & 1.038 & 1.201 & 1.395 & 0.015\\
Interest in AI & 0.095 & 0.046 & 1.005 & 1.099 & 1.203 & 0.039\\
\bottomrule
\multicolumn{7}{l}{\rule{0pt}{1em}\textit{Note: }}\\
\multicolumn{7}{l}{\rule{0pt}{1em}$N$=5,462, Nagelkerke $R^2$=.008, Model $\chi^2$(5)=23.6, p<.01}\\
\end{tabular}
\end{table}

The model reaches significance, but the independent variables only
explain 0.8\% of the variance of the dependent variable. Even though we find
significant effects for age, socioeconomic status, educational level,
and interest in AI, these effects are not substantial. The low salience of
ethical principles does not allow us to look for structural differences
in the population. This could change should the topic become more
prominent. Then it could be precisely these characteristics that account
for greater attention. At this point, however, such an assumption does
not hold up. Regarding RQ2, we found no notable effect of
educational level, socioeconomic status, and interest in AI on the
salience of ethical AI.

To answer H2, we calculated the co-occurrences of the different
categories of AI issue salience. Co-occurrences can only be detected if
participants mentioned at least two issues they were concerned with AI.
For each group we summed up the number of respondents who
mentioned issues of both respective categories.

AI ethics were mentioned 62 (6.6\%) times together with AI
functionalities, 218 (23.1\%) times together with AI applications and
202 (21.4\%) times together with other AI issues. However, 535 (56.7\%)
respondents mentioned AI ethics without any other category. In
comparison, AI applications (3,293, 79.9\%) were oftentimes named without
any other category. AI functionalities (448, 49.4\%) and AI issues (745,
58.5\%), on the other hand, were mentioned to an equal amount with other
categories. Though AI ethics are most commonly mentioned without any
other categories, it is mostly connected to specific AI applications or
other AI issues. Thereby, the huge number of single mentions is
connected to the fact that a majority of the respondents only were
concerned with AI in one issue (66.2\%) and did not indicate up to three
issues. Given that salience of ethical AI had the highest co-occurrence
with AI applications, H2 can be supported.

Lastly, we test if the different AI-related issues people were concerned
with have an impact on 1) intended AI avoidance (H3a) and 2) intention
to engage in public discussions about AI (H3b). For that, we calculated
two linear regression models with the dependent variables 1) intended AI
avoidance and 2) intention to publicly speak about AI with the
independent variable salience of ethical AI issues. We included the salience
of AI applications, AI functionalities, and other AI issues as control
variables. We further control the effects with the variables used in the
models before---age, gender, educational level, socioeconomic status,
and AI interest. We calculated the model for the respondents, who were
concerned with AI omitting those, who are disengaged with AI. Thus, we
can investigate in detail which effects different categories of issue
salience of AI have. Tables 3 and 4 show the results of the linear
regression model.

\begin{table}

\caption{\label{tab:unnamed-chunk-6}OLS Regression on Avoidance Intention}
\centering
\begin{tabular}[t]{lllll}
\toprule
  & b & SE & $\beta$ & p\\
\midrule
Intercept & 2.703 & 0.125 &  & 0.000\\
Age & 0.008 & 0.001 & 0.109 & 0.000\\
SE Status (1=Below Median) & 0.195 & 0.031 & 0.083 & 0.000\\
Gender (1=Female) & 0.130 & 0.030 & 0.057 & 0.000\\
Educational Level & -0.052 & 0.028 & -0.026 & 0.061\\
Interest in AI & -0.336 & 0.017 & -0.256 & 0.000\\
Salience of Ethical AI & 0.310 & 0.046 & 0.098 & 0.000\\
Salience of AI Functionalities & -0.056 & 0.045 & -0.018 & 0.210\\
Salience of AI Applications & -0.135 & 0.040 & -0.057 & 0.001\\
Salience of of other AI Issues & 0.006 & 0.042 & 0.002 & 0.886\\
\bottomrule
\multicolumn{5}{l}{\rule{0pt}{1em}\textit{Note: }}\\
\multicolumn{5}{l}{\rule{0pt}{1em}$N$=5,421, Adj. $R^2$=.114, $F$(9,5411)=78.57, p<.01}\\
\end{tabular}
\end{table}

\begin{table}

\caption{\label{tab:unnamed-chunk-7}OLS Regression on Intention to Engage in Public Discussions}
\centering
\begin{tabular}[t]{lllll}
\toprule
  & b & SE & $\beta$ & p\\
\midrule
Intercept & 1.930 & 0.132 &  & 0.000\\
Age & -0.004 & 0.001 & -0.052 & 0.000\\
SE Status (1=Below Median) & -0.014 & 0.033 & -0.005 & 0.674\\
Gender (1=Female) & -0.178 & 0.032 & -0.073 & 0.000\\
Educational Level & -0.015 & 0.029 & -0.007 & 0.616\\
Interest in AI & 0.521 & 0.018 & 0.369 & 0.000\\
Salience of Ethical AI & 0.097 & 0.049 & 0.029 & 0.046\\
Salience of AI Functionalities & 0.077 & 0.048 & 0.022 & 0.107\\
Salience of AI Applications & 0.057 & 0.043 & 0.022 & 0.181\\
Salience of of other AI Issues & 0.066 & 0.045 & 0.022 & 0.143\\
\bottomrule
\multicolumn{5}{l}{\rule{0pt}{1em}\textit{Note: }}\\
\multicolumn{5}{l}{\rule{0pt}{1em}$N$=5,357, Adj. $R^2$=.156, $F$(9,5347)=110.8, p<.01}\\
\end{tabular}
\end{table}

Concerning the OLS regression on avoidance intention, the independent
variables explain 11.4\% of the variance for the dependent variable
avoidance intention. Among the issue salience variables, ethical
salience had the strongest impact on avoidance intention. Avoidance of
AI rather occurs if people have ethical issues of AI top of their mind,
supporting H3a. Conversely, people who are concerned with specific
applications show fewer intentions to avoid AI. The salience of AI
functionalities and other AI issues do not have a significant effect on
avoidance behavior. Concerning the control variables, AI interest is the
strongest negative predictor of the dependent variable. We also find
significant effects for age, socioeconomic status, and gender. Among
people who are concerned with AI, those who are more interested in AI
show fewer intentions to avoid AI. Additionally, older people, people
with a socioeconomic status below average, and women report higher
avoidance tendencies.

Regarding the OLS regression on the intention to engage in public
discussion about AI, the model explains 15.6\% of the variance of the
dependent variable. Hereby, only the salience of ethical aspects of AI has a
significant, yet small, effect on the intention to speak about AI. Those
who are concerned with ethical topics are more willing to talk about AI.
Thus, H3b can be accepted. Respondents who have other AI topics in mind
show no significant tendency to talk more or less about AI. Again, we
find the strongest effect for interest in AI, followed by gender and age
as significant predictors. Higher interest in AI leads to a higher
intention to engage in public discussions about AI. This tendency is
stronger for men and younger people.

\hypertarget{discussion}{%
\section{Discussion}\label{discussion}}

Our study sheds light on a rather understudied perspective on ethical AI---public opinion. We investigated this perspective by focusing on
the salience of AI issues. For that, we collected data from 14,988 German
respondents.

\hypertarget{low-salience-of-ethical-issues-and-its-implication-for-implementing-ai-ethics}{%
\subsection{Low Salience of Ethical Issues and its Implication for
Implementing AI
Ethics}\label{low-salience-of-ethical-issues-and-its-implication-for-implementing-ai-ethics}}

Our results show that AI does not play a major role in the minds of
German citizens with about three-fifths not been concerned with AI
recently. However, when thinking about AI it is foremost thinking about
specific applications, especially autonomous driving. Ethical issues
only play a subordinate role among the German public, and if so, the
major share does not think about issues of fairness, accountability, and
transparency to which many researchers in the field of Fair Machine
Learning are devoted. These results indicate that the scientific---and
partially political---discourse does not reach the broad public. Ethical
issues of AI are mainly a topic, which is discussed by researchers.
However, leaving out respectively not reaching the public may interfere
with the normative goal of ethical development for the Common Good.
Without a broad public discourse about these topics, it is very unlikely
that public demands for the development of ethical AI will emerge
\citep{Kieslich.2021c}. It is not plausible that political institutions
or companies invest massive amounts of financial and personnel resources
in the development of ethical AI that benefits the whole society and
treats people equally if the public does not demand it. At this point,
it seems more likely that political institutions set up guidelines but
make compromises to hold their own in global competition
\citep{Hagendorff.2020}. Likewise, as long as companies make vast
profits and do not provoke massive scandals, economic actors presumably
will develop AI systems in a way that satisfies their stakeholders
\citep{Hagendorff.2020}. Importantly, only addressing stakeholders
leaves out a significant portion of society.

\hypertarget{how-likely-is-critical-engagement-with-ai}{%
\subsection{How Likely is Critical Engagement with
AI?}\label{how-likely-is-critical-engagement-with-ai}}

This leads to a reflection on the future of society---and who is
affected by the ethical threats of AI. Like the study of \citet{Frey.2017}
show for the context of workplace development, those who have lower
income and educational levels are most susceptible to negative
consequences of automation. Alarmingly, we observe that it is exactly
those people who do not care much about AI. Educational level is---besides
interest in AI---a strong predictor of being concerned with AI. Thus,
those who will be negatively affected the most are not concerned with
it. However, we did not find effects for gender and higher socioeconomic
status for being concerned with AI. As ethical issues with AI often
relate to women being discriminated against \citep{Tambe.2019} or lower incomes
resulting in deprivation by algorithms \citep{Pandey.2021}, it is also
noteworthy that these groups of people are not particularly concerned
with AI issues, which would play into the hands of economic and
political actors as previously argued. Our data show that greater
attention to ethical issues regarding AI is linked to both the intention
to avoid AI and the intention to participate more intensively in
debates. As a result, those who pay attention to ethical issues
regarding AI---which in most cases are potential breaches of ethical
standards---do not contribute to greater corporate profits and may even
steer a discourse on AI toward problematic aspects. If people are more
concerned with the ethical challenges AI systems pose for society, a
more critical way of citizen engagement with AI will seem possible.
Instead, the current discourse tends to result in a strongly product- and
application-oriented way of thinking about AI. Thus, the public AI
agenda reflects quite well the commerce-oriented tone in the media
discourse about AI. Having the high amount of mentioned AI applications
in contrast to ethical issues in mind, we describe the current state of
critical engagement with AI in Germany as low. Yet, the results suggest
that, if ethical issues are made prevalent, citizens are more reluctant
to integrate AI in their lives or to publicly engage with AI.

In summary, ethical issues of AI are not very prevalent among the German
public. This may be due to the lack of a critical use case, which fueled
public outcry against an AI system in Germany. Different from countries
such as the UK \citep{Kelly.2021} or France \citep{Wenzelburger.2021},
there has not been a major scandal in Germany concerning AI yet.
Additionally, we attribute these results to the low level of media
coverage on ethical aspects of AI \citep{Fischer.2021} and the dominance
of actors from the economic or political sector in news reporting.
Although there are some examples where specific AI applications were
blocked due to public awareness \citep{AlgorithmWatch.2020}, we are far
away from a broad discourse about the ethical challenges of AI systems. Yet,
we also detected in our data, that ethical issues may have an impact on
citizens' behavior towards AI. We also found that
ethical issues may have an impact on citizens' behavior towards AI when
such issues are prevalent in the minds of citizens. Future studies
should monitor the ongoing public debate about ethical AI and
investigate, whether certain key events or a possible change towards 
more critical news reporting will affect the broad public opinion on AI.
At least, for AI for the Common Good, it is relevant that the public is
involved in an ethical debate about AI.

\hypertarget{limitations}{%
\section{Limitations}\label{limitations}}

Our study has several limitations that need to be acknowledged. First of
all, we conducted our study in Germany. Citizens of other countries
might have different concerns with AI, for instance, \citet{Kelley.2021}
showed that public opinion on AI differed largely between citizens of
different countries. In the case of the focus on public concerns about AI
ethics, it would be especially interesting to compare our results with
those of UK or French citizens, where public debates about ethical issues
of AI have taken place lately \citep{Kelly.2021, Wenzelburger.2021}. It
would be interesting to investigate, whether a scandal might have an
influence on public concerns with ethical issues of AI in the long term.
Accordingly, we encourage researchers to replicate our survey in other
countries.

We also point out some potential issues with our measurement of AI
issues. We deliberately chose a broad approach with an open-ended
question to grasp public concerns with AI. However, we had to reduce
the complexity of our coding process. Thus, if two codes were applicable
for one answer, we had to choose one code, which was in focus. However,
some information might have gotten lost during this process. Moreover, not
being concerned with AI, and especially ethical issues, does not
necessarily have to mean that the public is not aware of these issues.
It means, however, that these issues are not of immediate concern and
arguably are not on the personal agenda. As we rely on agenda-setting
theory, the measurement chosen most adequately grasps our concept.
Finally, our measurement is not fine-grained enough to go into detail
about the concerns the public has. For example, it might very well be that
respondents, who answered ``autonomous driving'' in the open-ended
question think of possible ethical problems (e.g.,~the trolley problem).
But if the answer was only ``autonomous driving'' it was coded as such
(with it being value-free). Though, if respondents answered ``ethical
problems with autonomous driving'', it was coded as an ``ethics'' issue.
As we aimed for an overview analyzing the public agenda of AI issues, we
believe our measurement is satisfactory. Further research should
nevertheless build on our measurement and extend it in a way that the
tonality of several issues is measured as well.

\hypertarget{implications}{%
\section{Implications}\label{implications}}

We empirically showed that ethical AI plays a minor role in the minds of
German citizens. Given the media reporting about AI in Western
countries, which highlights economical aspects and oftentimes neglects
ethical issues, this finding is not surprising. Though, media outlets
oftentimes are self-committed to reporting in a multi-faceted way about
issues that are of importance to citizens. Although
\citet{Ouchchy.2020} and \citet{Chuan.2019} found that ethical issues
are on the rise, news reporting on AI does not seem to be very balanced
yet. Including ethical aspects in news reporting could lead to a higher
salience of ethical issues on behalf of media consumers. This could,
for example, be achieved by picking up ethical topics or including more
perspectives of researchers or NGOs, who are actively engaging with the
implementation of ethical AI. Currently, we see the dominance of
economic and political goals mirrored in the public opinion towards AI
in Germany. Vividly, German citizens are especially concerned with AI in
the contexts of autonomous driving or advances in medicine, which are
applications that the industry is vigorously pushing forward.

Though, increasing attention on the media side is not the only way to
fuel the salience of ethical AI issues. \citet{Long.2020} define AI
competencies that are needed for citizens to deal with AI reasonably.
Beneath technical competencies, they argue that
awareness of ethical AI should be taught to citizens. For that, literacy
projects are needed that expand the current state of AI literacy
programs that mostly focus on technical aspects of AI. However, for a
curriculum in the sense of AI for the Common Good, societal and ethical
issues should be included. Several programs already
offer a wider scope on AI, for example, Elements of AI
\citep{MinnaLearn.2022}. In the long term, it also could be useful to
integrate the topic of AI ethics into school or university curricula.
Thus, students who study computer science would also learn about the
(unintended) ethical risks of the technology and implement this
knowledge in their further development of AI systems.

Additionally, strengthening civil society actors could be a promising
way to fuel the public debate about ethical AI. Civil society actors
could actively address journalists, so that media coverage includes
positions of the civil society regularly. It was also shown
that some civil society groups exerted public pressure in some use cases
\citep{AlgorithmWatch.2020}. However, the inclusion of civil society is
rather an exception. Strengthening civil society can be achieved by
investing in human and material resources. This is all the more
important because civil society actors or academics advocating ethical
AI contrast with economic or political actors who have many times more
resources at their disposal. For a more balanced public discourse, there
needs to be at least a somewhat more even balance between the different
stakeholder groups.

We conclude that the salience of ethical AI issues is very low. For AI to
benefit the Common Good, the public debate about this topic needs to be
strengthened and also carried out to those who are affected by negative
effects of unethical AI. We find that there indeed is potential that
citizens can critically engage with AI; however, it needs to be
activated by media, civil society actors, researchers, and other
stakeholders who aim for AI for the Common Good.

\hypertarget{funding}{%
\section*{Funding}\label{funding}}
\addcontentsline{toc}{section}{Funding}

This study was conducted as part of the project \emph{Meinungsmonitor
Künstliche Intelligenz} (opinion monitor artificial intelligence). From
January 2020 to March 2021 the project was funded by the
\emph{Ministerium für Kultur und Wissenschaft des Landes
Nordrhein-Westfalen} (Ministry of Culture and Science of the State of
North Rhine-Westphalia), Germany. Since April 2021 the project is funded
by the \emph{Stiftung Mercator}.

\bibliographystyle{apacite}  
\bibliography{references}  

\hypertarget{appendix}{%
\section*{Appendix}\label{appendix}}
\addcontentsline{toc}{section}{Appendix}

\setcounter{table}{0}
\renewcommand{\thetable}{A\arabic{table}}

\begin{table}[H]

\caption{\label{tab:unnamed-chunk-8}Complete List of AI Issues}
\centering{
\begin{tabular}[t]{l>{\raggedright\arraybackslash}p{35em}}
\toprule
Category & Codes (Frequency)\\
\midrule
Functionalities & Robotics (266), Recognition (229), Algorithm (88), Chatbots (77), Machine Learning (74), Learning (64), Deep Learning (37), Prediction (25), Decision-Making (18), Recommendation (12), Robotic Process Automation (8), Neuromorphic Computing (7), Natural Language Processing (4)\\
Applications & Mobility (2087), Medicine (997), Smart Home (587), Communication (362), Industry (344), Entertainment (101), Military (98), Science (96), Finance (91), Service Tasks (64), Education (57), Climate (37), Human Resources (26), Public Administration (24), Law Enforcement (22), Agriculture (17), Politics (12), Law (9), Infrastructure (6)\\
Ethics & Control (262), Surveillance(185), Ethics (122), Manipulation (93), Security (70), Benevolence (42), Accountability (39), Transparency (32), Discrimination (26), Fairness (5)\\
Other Issues & Workplace (456), Digitalization (196), Future (97), Human-Machine-Interaction (72), Media (48), Economical Growth (23), Investment (19), Overload (12), Governance (6)\\
\bottomrule
\end{tabular}}
\end{table}

\end{document}